\documentclass{article}

\usepackage{mathpple}
\usepackage{graphicx}
\usepackage{url}
\usepackage{paralist}

\usepackage{geometry}
\begin{document}

\noindent
\textbf{Preprint of:}\\
T. A. Nieminen, V. L. Y. Loke,
A. M. Bra\'{n}czyk,
N. R. Heckenberg and
H. Rubinsztein-Dunlop\\
``Towards efficient modelling of optical micromanipulation\\
of complex structures''\\
pp. 442--446 in
\textit{PIERS 2006 Cambridge Proceedings},
The Electromagnetics Academy, Cambridge, MA (2006)

\hrulefill

\begin{center}

\Large
\textbf{Towards efficient modelling of optical micromanipulation\\
of complex structures}

\normalsize
T. A. Nieminen, V. L. Y. Loke,
A. M. Bra\'{n}czyk,
N. R. Heckenberg and
H. Rubinsztein-Dunlop

\textit{The University of Queensland,
Brisbane, Australia}

\texttt{timo@physics.uq.edu.au}

\end{center}

\begin{abstract}
Computational methods for electromagnetic and light
scattering can be used for the calculation of optical
forces and torques. Since typical particles that are
optically trapped or manipulated are on the order of the
wavelength in size, approximate methods such as geometric
optics or Rayleigh scattering are inapplicable, and solution
or either the Maxwell equations or the vector Helmholtz equation
must be resorted to. Traditionally, such solutions were only
feasible for the simplest geometries; modern computational
power enable the rapid solution of more general---but still
simple---geometries such as axisymmetric, homogeneous, and
isotropic scatterers. However, optically-driven micromachines
necessarily require more complex geometries, and their
computational modelling thus remains in the realm of challenging
computational problems. We review our progress towards efficient
computational modelling of optical tweezers and micromanipulation,
including the trapping and manipulation of complex structures
such as optical micromachines. In particular,
we consider the exploitation of symmetry in the modelling
of such devices.
\end{abstract}

\section{Introduction}

Optical tweezers have seen deployment in a wide range of applications
in biology, soft materials, microassembly, and other fields.
As well as being used
for the trapping and manipulation of a wide range of natural and artificial
objects, optically trapped probes are used to measure forces on the order
of piconewtons. Compared with this diverse range of experimental applications,
theory and accurate computational modelling of optical tweezers has received
much less attention and has remained relatively undeveloped, especially for
non-spherical particles and non-Gaussian beams. This is unfortunate,
especially when we consider the growing
fields of controlled rotation of complex microparticles---prototype
optically-driven micromachines---and fully three-dimensional
manipulation using complex optical fields, where
the application of theory and modelling provide insight into the
physics, and allow engineering and optimisation.

Since optical forces and torques result
from the transfer of momentum and angular momentum from the trapping
beam to the particle via scattering, the theory and computational
modelling of optical tweezers is, in essence, the theory and computational
modelling of the scattering of light or electromagnetic radiation.
Since typical particles that are
optically trapped or manipulated are on the order of the
wavelength in size, approximate methods such as geometric
optics or Rayleigh scattering are inapplicable, and solution
or either the Maxwell equations or the vector Helmholtz equation
must be resorted to. As scattering by particles in this size
range is of interest in many fields, a wide variety of analytical
and computational methods have been developed. Thus, there is
a solid foundation on which to develop computational modelling
of optical micromanipulation.

There are, however, complications that prevent simple direct application of
typical light-scattering codes. The first, but not necessarily the
most important, is that optical tweezers
makes use of a highly focussed laser beam, while most existing scattering
codes assume plane wave illumination. Perhaps more fundamental is the
need for a large number of repeated calculations to characterise an
optical trap---even for an axisymmetric (but nonspherical) particle trapped
in a circularly polarised Gaussian beam, we already have four degrees
of freedom. Clearly, this places strong demands on computational
efficiency.

Due to this requirement for repeated calculation of scattering by
the same particle, we employ the \textit{T}-matrix
method~\cite{waterman1971,mishchenko2004b}.
Below, we outline the employment of the \textit{T}-matrix method
for the calculation of optical forces and torques. While most
implementations of the \textit{T}-matrix method are
restricted to simple geometries, this is not a limitation
inherent in the method; fundamentally, the \textit{T}-matrix
method is a \emph{description} of the scattering properties of
a particle, not a method of calculating the scattering properties.
Therefore, in principle, any method of calculating scattering
can be used to obtain the \textit{T}-matrix for a scatterer.
We discuss such ``hybrid'' methods, where a computational method
not usually associated with the \textit{T}-matrix method is
used to calculate the \textit{T}-matrix of a scatterer, and hence
the optical force and torque.

A further important consideration is that optical micromachines,
while complex, are likely to possess a high degree of symmetry; this
can be exploited to reduce computation times by orders of magnitude.
We demonstrate the effectiveness of this approach by modelling
the optical trapping and rotation of a cube. The two principal
symmetries of such shapes---mirror symmetry and discrete rotational
symmetry about the normal to the mirror symmetry plane---are
exactly the symmetries that typify the ideal optically-driven rotor.

\section{\textit{T}-matrix formalism for optical force and torque}

The \textit{T}-matrix method in wave scattering involves writing the
relationship between the wave incident upon a scatterer, expanded in terms of
a sufficiently complete basis set of functions $\psi_n^{(\mathrm{inc})}$,
where $n$ is a mode index labelling the functions, each of which is a
solution of the Helmholtz equation,
\begin{equation}
U_\mathrm{inc} = \sum_n^\infty a_n \psi_n^{(\mathrm{inc})},
\label{exp1}
\end{equation}
where $a_n$ are the expansion coefficients for the incident wave,
and the scattered wave, also expanded in terms of a basis set
$\psi_k^{(\mathrm{scat})}$,
\begin{equation}
U_\mathrm{scat} = \sum_k^\infty p_k \psi_k^{(\mathrm{scat})},
\label{exp2}
\end{equation}
where $p_k$ are the expansion coefficients for the scattered wave,
is written as a simple matrix equation
\begin{equation}
p_k = \sum_n^\infty T_{kn} a_n
\end{equation}
or,  in more concise notation,
\begin{equation}
\mathbf{P} = \mathbf{T} \mathbf{A}
\end{equation}
where $T_{kn}$ are the elements of the \textit{T}-matrix. The
\textit{T}-matrix formalism is a Hilbert basis description of
scattering. The \textit{T}-matrix depends only on the properties of the
particle---its composition,  size,
shape, and orientation---and the wavelength, and is otherwise
independent of the incident field.

This means
that for any particular particle, the \textit{T}-matrix only needs to be
calculated once, and can then be used for repeated calculations.
This is the key point that makes this an attractive method for modelling
optical tweezers, providing a
significant advantage over many other methods of calculating scattering where
the entire calculation needs to be repeated.

The natural choice of basis functions when describing scattering
by a compact particle is to use vector spherical wavefunctions
(VSWFs)~\cite{waterman1971}.
The optical force and torque are given by sums of products of the modal
amplitudes~\cite{farsund1996,crichton2000,nieminen2004d}.

Notably, neither how the VSWF expansion of the incident field nor how
the \textit{T}-matrix can be calculated has entered the above description
of scattering.
A variety of methods exist for the
former~\cite{nieminen2003a,nieminen2004d}, and the latter task is generally
the more challenging computationally.

Most implementations of the \textit{T}-matrix method use the extended
boundary condition method (EBCM), also called the null field method, to
calculate the \textit{T}-matrix. This is so widespread that the
\textit{T}-matrix method and the EBCM are sometimes considered to be
inseparable, and the terms are sometimes used interchangeably.
However, from the description above, it is clear that the \textit{T}-matrix
formalism is independent of the actual method used to calculate the
\textit{T}-matrix~\cite{kahnert2003b,nieminen2003b}.

A number of alternative methods have been used for the calculation of
\textit{T}-matrices.
Notably, such ``hybrid'' methods, for example the discrete dipole
approximation (DDA) method used by
Mackowski~\cite{mackowski2002} can be used for the calculation of
\textit{T}-matrices for particles of arbitrary shape, internal structure,
and electromagnetic properties. Complex internal structure will generally
require a discretisation of the internal volume of the particle, rather
than a method based on surface discretisation. We are working on
both finite-difference frequency-domain (FDFD) and DDA
based hybrid \textit{T}-matrix solvers.

\section{Optical torque and symmetry}

The \textit{T}-matrix elements are strongly dependent on the symmetry
of the scatterer~\cite{waterman1971}. We can deduce the principal features
from Floquet's theorem, relating solutions to differential equations to
the periodicity of their boundary conditions.

If we have a scatterer with $n$th-order rotational symmetry about the
$z$-axis, an incident mode of azimuthal index $m$ couples to scattered
modes with azimuthal indices $m, m\pm n, m\pm 2n, m\pm 3n$ and so on. For
scatterers that are mirror-symmetric, upward and downward coupling must be
equal, in the sense that, for example, a mirror-symmetric scatterer of
2nd order rotational symmetry (such as a long rod), {T}-matrix elements
coupling from $m = 1$ modes to $m = -1$ modes will have the same magnitudes
as the elements coupling from $m = -1$ to $m = 1$ modes. For chiral
scatterers, these \textit{T}-matrix elements will, in general, be different.

This directly affects the optical torque; the vector spherical wavefunctions
are eigenfunctions of the angular momentum operators $J^2$ and $J_z$.
Essentially, the radial mode index $n$ gives the magnitude of the
angular momentum flux, while the azimuthal mode index $m$ gives
the $z$-component of the angular momentum flux. Therefore, the coupling
between orders of different $m$ describes the generation of optical
torques about the beam axis.

For the case of a rotationally symmetric scatterer, this means that there
is no coupling between modes with differing angular momenta about
the $z$-axis\cite{waterman1971,mishchenko1991,nieminen2004a}.
Therefore, it is not possible to exert optical torque on such scatterers
except by absorption (or gain)---since the incoming and outgoing
angular momenta
per photon are the same, the only optical torque can result from a change
in the number of photons. In general, the use of absorption for the
transfer of optical torque is impractical, due to excessive heating.
Therefore, a departure from rotational symmetry is required. This
can be either at the macroscopic (the shape of the particle) or
microscopic (optical properties of the particle) level.

Birefringent and elongated or flattened particles are simple
examples of introducing such asymmetry; notably, such particles were
the first to be controllably optically rotated through means other
than absorption, for example by Beth in the first measurements
of optical torque~\cite{beth1936}. Particles with these properties
have also been rotated in optical traps~\cite{friese1998nature,%
bonin2002,bayoudh2003,bishop2003}. As such particles can still
be axisymmetric about one axis, rapid calculation of optical
forces and torques is still possible~\cite{bayoudh2003,bishop2003}

More complex particles have also been fabricated and
rotated~\cite{galajda2001,luo2000,ukita2002}, but in these cases,
there are few results from computational modelling~\cite{collett2003}.

As such structures typically possess discrete rotational symmetry,
the restrictions on coupling between azimuthal orders can be used
to reduce the number of \textit{T}-matrix elements that need to
be calculated. This can greatly reduce the time required. This
is also the case for the hybrid methods described above. For a scatterer
with $p$th-order discrete rotational symmetry, it is only necessary
to perform calculations for a $1/p$ portion of the entire structure.
If, in addition, there is mirror symmetry about the $xy$ plane, the
parity of the VSWFs will be preserved. Therefore, an odd-$n$ TE mode
will only couple to odd-$n$ TE modes and even-$n$ TM modes. This halves
number of non-zero \textit{T}-matrix elements, and halves the portion
of the structure that needs to be modelled.

\section{Example: optical trapping of a cube}

A simple example illustrating both the relationship between optical
torque and symmetry, and the exploitation of particle symmetry for
more efficient calculation of optical forces and torques,
is the optical trapping of a cube. The cube embodies both
of the symmetries---mirror symmetry and discrete rotational
symmetry about the normal to the mirror symmetry plane---that
typify the ideal optically-driven rotor.

As the cube has 4th-order rotational symmetry, and mirror symmetry
with respect to the $Cy$ plane, each incident modes only couples to
approximately $1/8$ the number of significant scattered modes.
Although the column-by-column calculation of the \textit{T}-matrix
still requires the same number of least-squared solutions, each of
this is of a smaller system of equations, and much faster.
For example, the two wavelengths wide cube used in our example below
required 30 minutes for the calculation of the \textit{T}-matrix
on a 32 bit single-processor 3\,GHz microcomputer, as compared with
30 hours for an object of the same size lacking the cube's symmetries.
Only one octant of the cube was explicitly included in the
calculation.

If figure 1, we show the optical force and torque exerted on a
cube with relative refractive index of $1.19 = 1.59/1.34$, and faces
$2\lambda$ across, where $\lambda$ is the wavelength in the surrounding
medium. Once the \textit{T}-matrix is calculated, to calculate the
optical force and torque at a particular position requires less than
1 second (unless the point is far from the beam focus, in which case,
up to 10 seconds or so can be needed).

\begin{figure}[!h]
(a) \hspace{0.47\columnwidth} (b) \\
\centerline{\includegraphics[width=0.48\columnwidth,draft=false]{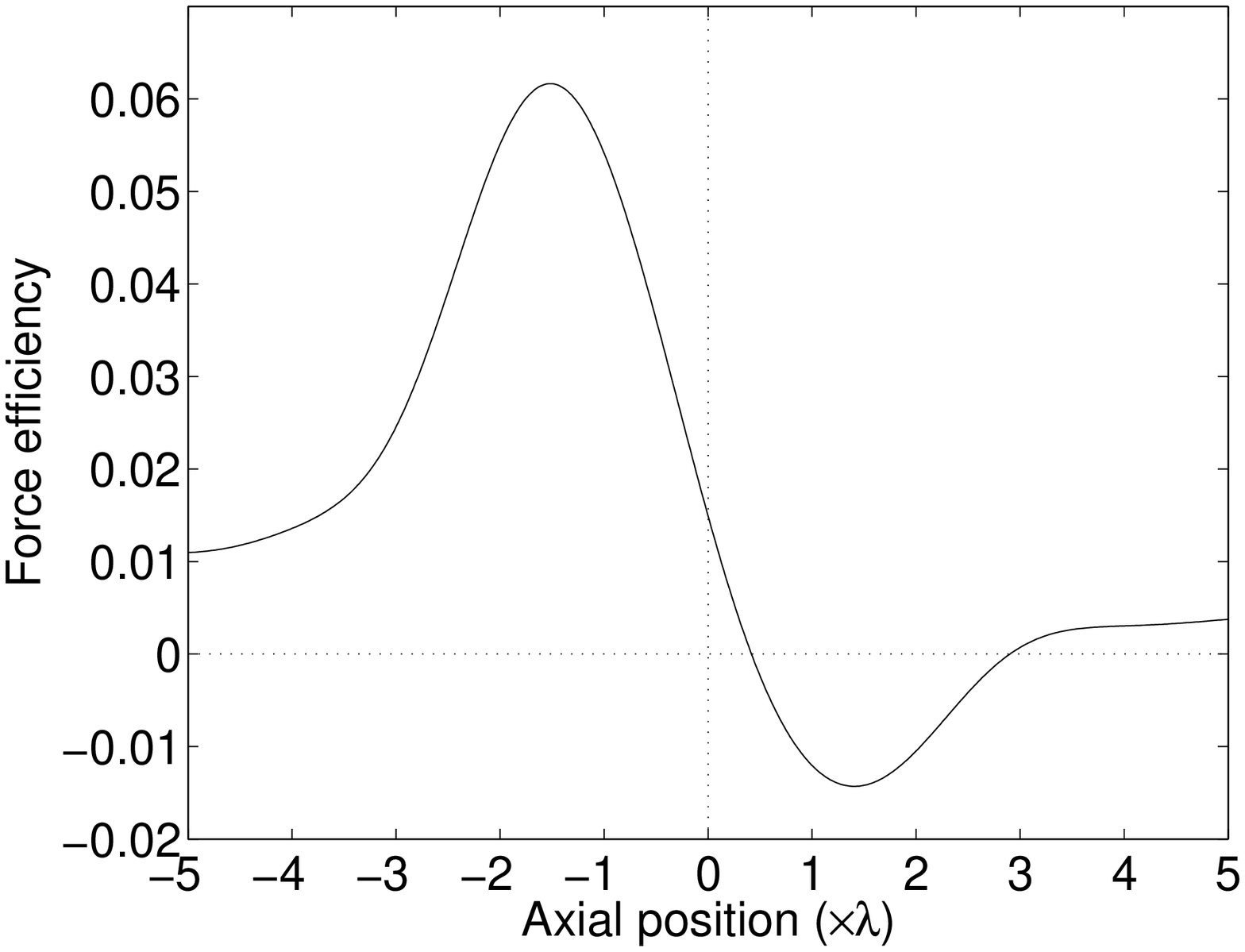}
\includegraphics[width=0.48\columnwidth,draft=false]{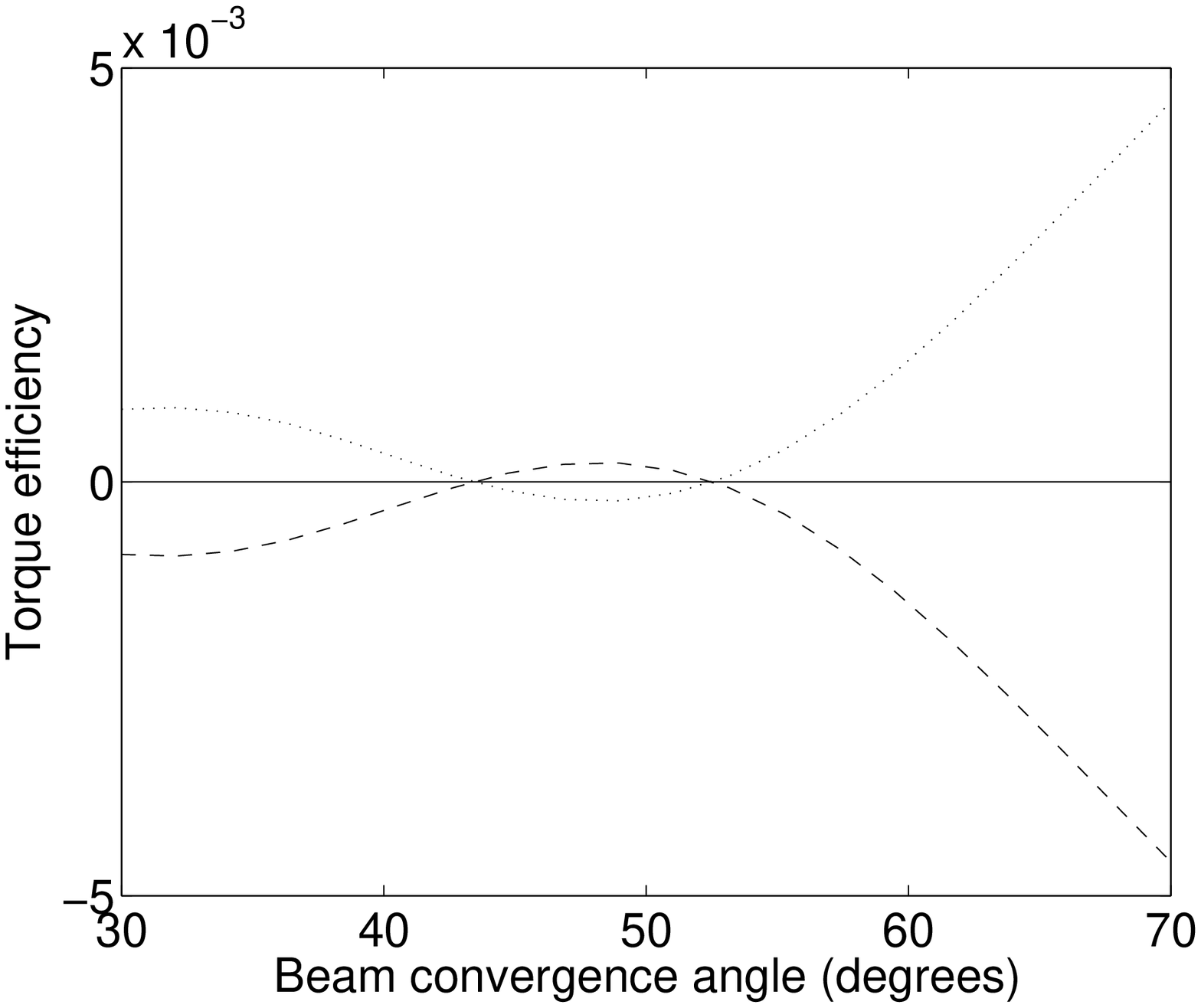}} \\
(c)  \hspace{0.47\columnwidth} (d) \\
\centerline{\includegraphics[width=0.48\columnwidth,draft=false]{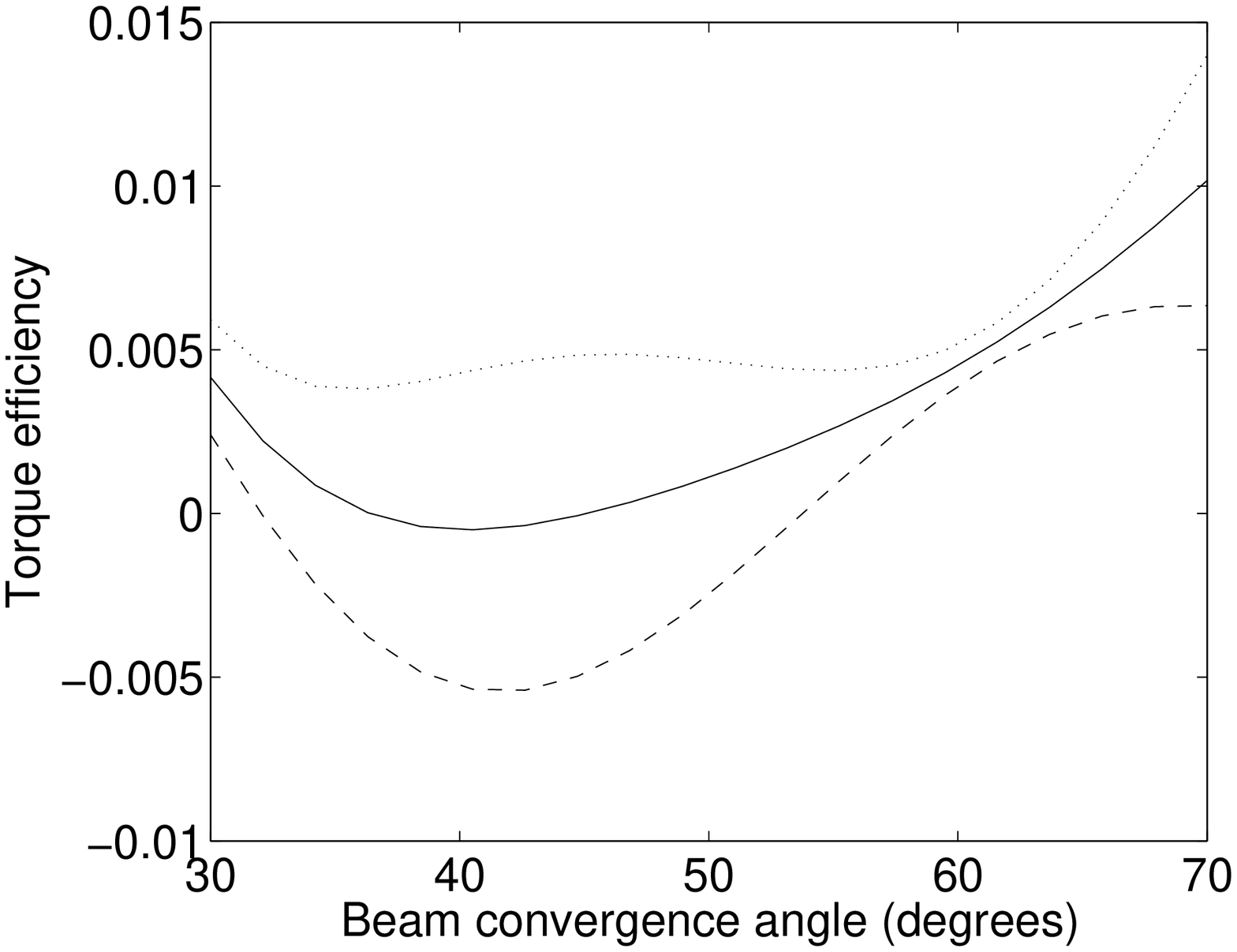}
\includegraphics[width=0.48\columnwidth,draft=false]{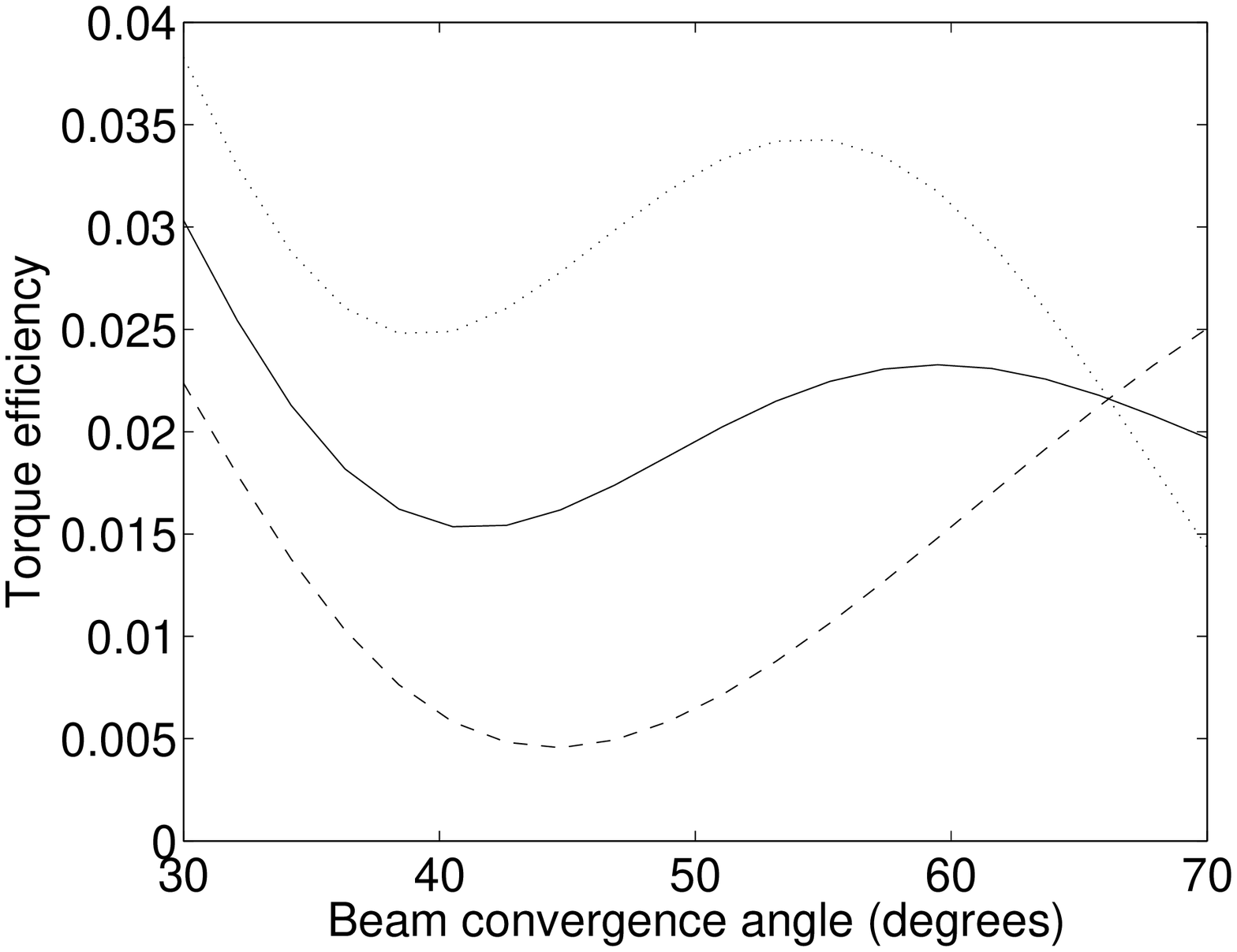}} 
\caption{Optical force and torque on a dielectric cube. (a) shows the
axial force as a function of position along the beam axis, showing that
the cube can be trapped. (b)--(d) show the dependence of the optical
torque on the beam convergence angle and the polarisation and orbital
angular momentum. In (b), the beam is Gaussian (ie LG$_{00}$), while in
(c) and (d), the beams are LG$_{01}$ and LG$_{02}$ respectively.
The solid lines are for plane polarised beams, while dotted and dashed
lines are for circularly polarised beams with spin parallel to and
antiparallel to the orbital angular momentum.}
\label{fig1}
\end{figure}

In figure 1(a), we see that cubic shapes can be stably trapped axially,
while 1(b)--(d) show that optical torque can be generated by such
structures. The increased efficiency resulting from the use of 
orbital angular momentum~\cite{nieminen2004d} is clear.

\section{Conclusion}

The symmetry properties of a scatterer can be used to dramatically speed
the calculation of the scattering properties of a particle. If these
are expressed in the form of the \textit{T}-matrix, this enables rapid and
efficient calculation of optical forces and torques. Since typical
optically-driven microrotors possess discrete rotational symmetry, they are
ideal candidates for this method. In addition, mirror symmetry about
a plane can also be used to further reduce the computational burden.
Finally, ``hybrid'' \textit{T}-matrix methods can be used for particles
with geometries or internal structure making them unsuitable for
traditional methods of calculating \textit{T}-matrices.

%\bibliography{../../journalsabbrev,../../papers,extra_refs}
%\bibliographystyle{ieeetr}   %>>>> makes bibtex use spiebib.bst

\end{document}